\documentclass[floatfix, aps,prb,preprint, groupedaddress]{revtex4-2}
\usepackage{dcolumn}
\usepackage{bm}
\usepackage{amssymb}
\usepackage{amsthm}
\usepackage{physics}
\usepackage{color}
\usepackage[version=4]{mhchem}
\usepackage{graphicx}
\usepackage{natbib}
\usepackage{subfig}
\usepackage{booktabs}
\usepackage{multirow}
\usepackage[T1]{fontenc}

\newcommand{\degC}{$^\circ$C}

\newcommand{\TC}{T_{\rm C}}

\newcommand{\uB}{\mu_{\rm B}}

\newcommand{\Hm}{H_{\rm m}}

\newcommand{\LIC}{\ce{LuInCo_4}}
\newcommand{\LC}{\ce{LuCo_2}}

\newcommand{\YC}{\ce{YCo_2}}

\newcommand{\bQ}{{\bm q}}
\begin{document}


\title{Gradual Localization of Itinerant Cobalt 3$d$ Electrons in the Pyrochlore Ferromagnetic Metal LuInCo$_4$}


\author{Taiki Shiotani}
\email[]{Corresponding author: shiotani@issp.u-tokyo.ac.jp}
\altaffiliation{Present address: Institute for Solid State Physics, The University of Tokyo, Kashiwa, Chiba 277-8581, Japan}
\affiliation{Department of Materials Science and Engineering, Kyoto University, Kyoto 606-8501, Japan}

\author{Subaru Yoshimoto}
\affiliation{Department of Materials Science and Engineering, Kyoto University, Kyoto 606-8501, Japan}

\author{Yoshikazu Tabata}
\affiliation{Department of Materials Science and Engineering, Kyoto University, Kyoto 606-8501, Japan}

\author{Yusuke Nambu}
\affiliation{Institute for Integrated Radiation and Nuclear Science, Kyoto University, Kumatori 590-0494, Japan}

\author{Katsuki Kinjo}
\affiliation{Institute of Multidisciplinary Research for Advanced Materials, Tohoku University, Sendai 980-8577, Japan}

\author{Kazuhiro Nawa}
\affiliation{Institute of Multidisciplinary Research for Advanced Materials, Tohoku University, Sendai 980-8577, Japan}

\author{Taku J. Sato}
\affiliation{Institute for Solid State Physics, The University of Tokyo, Kashiwa, Chiba 277-8581, Japan}
\affiliation{Institute of Multidisciplinary Research for Advanced Materials, Tohoku University, Sendai 980-8577, Japan}

\author{Takeshi Waki}
\altaffiliation{Present address: Department of Molecular Chemistry and Biochemistry, Doshisha University, Kyotanabe 610-0321, Japan}
\affiliation{Department of Materials Science and Engineering, Kyoto University, Kyoto 606-8501, Japan}

\author{Hiroyuki Nakamura}
\affiliation{Department of Materials Science and Engineering, Kyoto University, Kyoto 606-8501, Japan}

\date{\today}

\begin{abstract}
We investigated the microscopic magnetic state of the itinerant-electron ferromagnet {\LIC} with a cobalt pyrochlore lattice by means of powder and single-crystal neutron diffraction measurements.
The powder neutron diffraction profiles at zero field are well described by a simple ${\bm q}=0$ collinear ferromagnetic structure with an ordered Co moment of $0.9\ \uB$.
Single-crystal neutron diffraction experiments further suggest the absence of both antiferromagnetic order and long-period helical magnetic modulation.
In addition, diffuse magnetic scattering was observed around ${\bm q}=0$, with its intensity distribution gradually broadened in reciprocal space upon cooling.
Based on these observations, we propose a possible microscopic origin of the low-temperature metamagnetic transition in terms of a ferromagnetic instability enhanced by gradual localization of the itinerant Co--3$d$ electrons.
These results suggest that {\LIC} provides a model system for investigating the interplay between electron localization and itinerant-electron ferromagnetism in cobalt-based metallic pyrochlore compounds.
\end{abstract}


\maketitle

\section{Introduction}
Characteristic lattice geometries have been a fertile ground for emergent quantum phenomena in condensed matter systems.
One such geometry is the pyrochlore lattice, consisting of a three-dimensional network of corner-sharing tetrahedra. This lattice has attracted considerable attention because its characteristic geometry strongly influences both localized and itinerant electrons.
In localized electron systems, geometrical magnetic frustration gives rise to exotic magnetic states, such as spin liquids, spin ice, and spin glasses \cite{spiniceDy,spiniceHo, spigla, QSL}.
In itinerant-electron systems, the pyrochlore geometry can create electronic structures with unconventional band topology. This can lead to anomalous transport phenomena, such as the anomalous and topological Hall effects \cite{Ir1, Ir2, Ir3}.
Metallic pyrochlore compounds, particularly the C15 Laves phases, have served as simple model systems exhibiting a remarkable variety of phenomena arising from the interplay between lattice geometry and itinerant electrons.
Examples include the frustration-driven long-period helical magnetic order in YMn$_2$ \cite{YMheli}, spin-liquid behavior in Sc-substituted YMn$_2$ \cite{QSLYMn}, three-dimensional flat bands in nearly ferromagnetic CaNi$_2$ \cite{CaNi}, and symmetry-protected topological band crossings in the superconductor {\ce{CeRu2}} \cite{CeRu}.
The cobalt-based compounds {\YC} and {\LC} are exchange-enhanced Pauli paramagnets \cite{Lemaire, Burzo, RC, RCpara}, in which the Co atoms do not carry spontaneous magnetic moments, although the 3$d$ bands exhibit a high density of states just below the Fermi level ($E_{\mathrm F}$) \cite{fd0, YCo2band1}.
This proximity to ferromagnetism gives rise to an itinerant-electron metamagnetic transition under high external magnetic fields \cite{YCo2metaSakaki,Ymeta, Lumeta}, while substitution with magnetic rare-earth ($R$) ions induces spontaneous magnetic ordering through the strong $R$-molecular field \cite{RC}.\\
\indent Recently, the cobalt-based, site-ordered C15b Laves phase compounds $RA$Co$_4$ ($A=$ Mg or In) were reported to exhibit magnetic properties different from those of the C15-type compounds \cite{YMC, LIC, TbMC, tsujii}.
YMgCo$_4$ and LuInCo$_4$ exhibit spontaneous ferromagnetism \cite{YMC,LIC}, which is induced by the nearly flat Co--3$d$ bands located close to the Fermi level.
Meanwhile, $R$InCo$_4$ ($R=$ Dy--Tm) compounds have been reported to exhibit room-temperature compensated ferrimagnetism with antiferromagnetic coupling between the rare-earth and Co sublattices \cite{RIC}.
Among this family, LuInCo$_4$ exhibits unusual ferromagnetic properties.
Although it undergoes a ferromagnetic transition at $\TC=306$ K, magnetic anisotropy with [001] as the easy magnetization axis develops below approximately 100 K, accompanied by a low-field metamagnetic transition when a magnetic field is applied along the hard [111] direction [Fig.\ \ref{gptas}(a)] \cite{LIC}.
These anomalous properties may be related to the geometrical effects of the pyrochlore Co sublattice, including the flat Co--3$d$ bands and possible frustration due to the local magnetic anisotropy.
However, the microscopic magnetic ground state and the origin of the metamagnetic transition remain unresolved.\\
\indent In this study, we investigate the magnetic ground state of LuInCo$_4$ by means of powder and single-crystal neutron diffraction measurements under a zero magnetic field.
The neutron diffraction results show no evidence of an antiferromagnetic component or long-period helical magnetic modulation, indicating that the magnetic structure is consistent with a collinear ferromagnetic state.
We also observed diffuse magnetic scattering around ${\bm q}=0$, and its intensity distribution gradually broadens in reciprocal space upon cooling. This suggests a gradual localization of the itinerant Co--3$d$ electrons.
Based on these findings, we propose a possible microscopic origin of the low-temperature metamagnetic transition in {\LIC} in terms of the residual ferromagnetic instability.
\\
\section{Experiments}
Single crystals of {\LIC} were grown using the self-flux method described in our previous report \cite{LIC}.
To obtain larger crystals, we modified the procedure. 
We started with Lu ingots (Nippon Yttrium, 99.9\% purity), In shots (Rare Metallic, 99.99\% purity), and Co flakes (Rare Metallic, 99.9\% purity).
The mixture with a composition of {\ce{LuIn_2Co_{2.5}}} was arc-melted in an argon atmosphere, followed by grinding. 
The resulting mixture was placed in a boron nitride (BN) crucible and sealed in a partially argon-filled molybdenum tube using arc welding.
The ampoule was then placed in a vertical-tube high-temperature furnace equipped with {\ce{MoSi_2}} heating elements (BF-1700-II, Crystal System Corp.). The {\ce{Al_2O_3}} ceramic furnace tube was evacuated during the synthesis to prevent oxidation of the Mo tube.
The ampoule was initially heated to 1400\ \degC\ over 7 h, held at that temperature for 0.5 h, cooled to 1150\ \degC\ over 8 h, slowly cooled to 975\ \degC\ at a rate of 3\ \degC$/$h, and finally furnace-cooled.
Octahedral or plate-like crystals with \{111\} faces were obtained.
The remaining flux and impurity phases on the surface were removed by immersion in dilute hydrochloric acid.
Crystal and magnetic structure refinements of the neutron diffraction data were performed using the FullProf Suite \cite{FullP1, FullP2}.
\\
\indent Powder neutron diffraction experiments were performed for {\LIC} using the neutron powder diffractometer HERMES installed at the JRR-3 reactor of the Japan Atomic Energy Agency (JAEA), Tokai, Japan \cite{Hermes}.
The neutron wavelength was $\lambda=1.34177$ \AA.
Diffraction data were collected over the scattering-angle range of $2\theta=2.317$--$152.668^\circ$ using 150 $^3$He detectors with a $2\theta$ step of $0.05^\circ$.
The powder sample was sealed in a cylindrical vanadium can containing $^4$He exchange gas to suppress unwanted Bragg reflections from the container itself. 
Measurements were carried out over the temperature range of $T=3$--400\ K. To this end, the powder was prepared by crushing single crystals.
\begin{figure}[h]
  \centering
  \includegraphics[keepaspectratio, width=0.85\columnwidth]{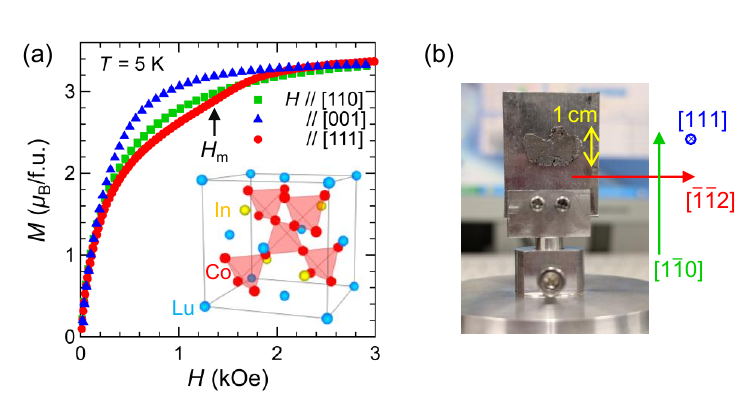}
  \caption{(a) Magnetic-field dependence of the magnetization at $T=5$\ K \cite{LIC}. The inset shows the crystal structure of {\LIC}. (b) Photograph of the single-crystal sample measured for the neutron diffraction experiments using the GPTAS (4G) triple-axis thermal neutron spectrometer.}
  \label{gptas}
\end{figure}
Single-crystal neutron diffraction measurements for {\LIC} were performed using the GPTAS (4G) triple-axis thermal neutron spectrometer installed at JRR-3 \cite{gptas}.
The spectrometer was operated in the triple-axis mode with horizontal collimations of $40'$-$40'$-$40'$-$80'$.
The incident neutron wavelength was $\lambda_i = 2.3665$\ \AA, and the wavelength was calibrated using an \ce{Al_2O_3} standard before the neutron diffraction measurements. To eliminate the higher harmonic neutron contribution, a PG filter was inserted upstream of both the PG monochromator and analyzer.
We used a large single crystal, with dimensions of $\sim15\times10\times0.87$\ mm$^3$, which was glued onto an aluminum plate with varnish and aligned with the $(hhl)$ plane as the scattering plane [Fig.\ \ref{gptas}(b)]. The sample was sealed in a standard aluminum sample can with $^4$He exchange gas. 
A $^4$He closed-cycle Gifford--McMahon refrigerator was used over the temperature range $T=3$--$300$ K.
\\
\section{Results}
\subsection{Powder neutron diffraction}
\begin{figure}[]
    \centering
    \includegraphics[keepaspectratio, width=\columnwidth]{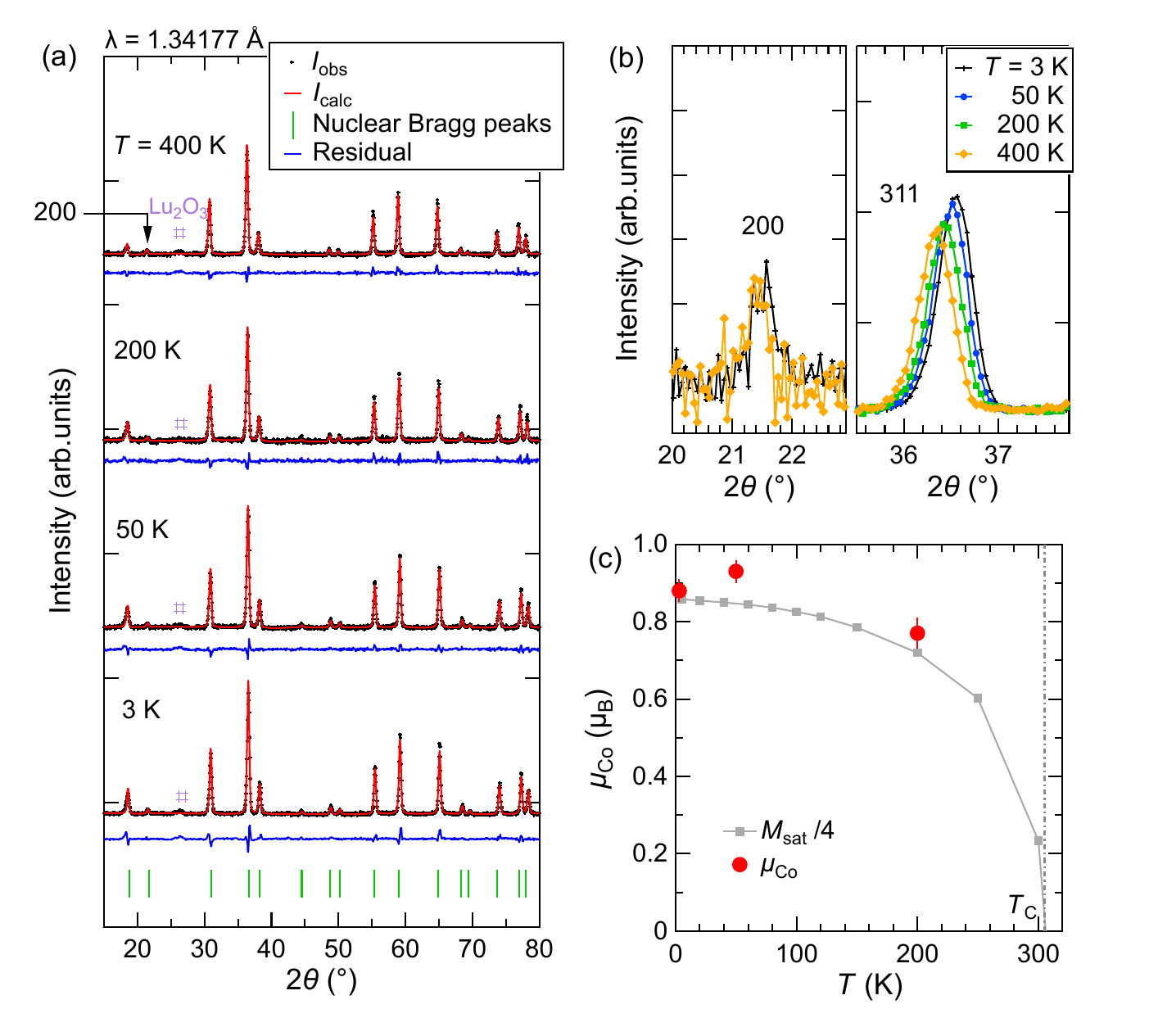}
\caption{(a) Powder neutron diffraction profiles of {\LIC} at different temperatures. The results of the Rietveld refinements, along with the expected Bragg reflection positions, are also shown. (b) Profiles in the extended 2$\theta$ range around the (200) and (311) Bragg reflections. (c) Temperature evolution of the refined magnetic moment at the Co site.}
    \label{pnd}
\end{figure}
\begin{table*}[]
\centering
\begin{ruledtabular}
\caption{Refined lattice parameters, atomic coordinates ($x, x, x$), and magnetic moments of the Co--16$e$ site. The Rietveld $R$ factor $R_{\rm wp}$, goodness-of-fit $\chi^2$, Bragg $R$ factor $R_{\rm Bragg}$, and magnetic $R$ factor
$R_{\rm mag}$ are also listed.}
\begin{tabular}{lccccccc}
$T$\ (K) & $a$\ (\AA) & $x$ & $\mu_{\rm Co}$\ ($\uB$) &
$R_{\rm wp}$ & $\chi^2$ & $R_{\rm Bragg}$ & $R_{\rm mag}$ \\
\hline
3   & 7.0111 & 0.6273 & 0.88(3) & 6.53 & 0.585 & 5.44 & 8.08 \\
50  & 7.0119 & 0.6269 & 0.93(3) & 7.35 & 0.659 & 6.41 & 10.3 \\
200 & 7.0229 & 0.6283 & 0.77(4) & 8.83 & 0.940 & 6.73 & 13.9 \\
400 & 7.0399 & 0.6245 & --      & 7.11 & 0.600 & 7.35 & --   \\
\end{tabular}
\label{nriet}
\end{ruledtabular}
\end{table*}
Figures\ \ref{pnd}(a) and \ref{pnd}(b) show the powder neutron diffraction (PND) profiles at different temperatures.
Upon cooling from $T = 400$\ K through the magnetic transition at $\TC = 306$\ K, the intensity of the peaks increases (see the right side of Fig.\ \ref{pnd}(b)), indicating the development of a magnetic contribution.
Meanwhile, the positions of the peaks shift to higher scattering angles, indicating a spontaneous volume contraction.
There is no observation of additional peaks at the structurally forbidden positions or peak splitting, suggesting that the magnetic and paramagnetic unit cells are the same.
For the Rietveld refinements, we assumed a simple collinear ferromagnetic spin structure.
For the present cubic powder diffraction data, the collinear ferromagnetic models with moments along [001], [110], and [111] yield indistinguishable calculated profiles. Therefore, the moment direction cannot be determined from the present PND data.
The results of the refinements using FullProf are shown in Fig.\ \ref{pnd}(a) and Table\ \ref{nriet}.
The estimated magnetic moment of the Co site at $T=3$\ K is $\mu_{\rm Co} = 0.88(3)$\ $\uB$. This value is similar to that obtained from DFT calculations (1.3\ $\uB$) and comparable to the saturation magnetization at $T = 5$\ K (3.43\ $\uB$/{f.u.} \cite{LIC}).
The value remains nearly unchanged between 3 and 50 K within the experimental uncertainty and decreases at 200 K [Fig.\ \ref{pnd}(c)], consistent with the temperature dependence of the saturation magnetization.
Previous DFT calculations have suggested that the Lu site carries a small magnetic moment that is antiparallel to the Co moments.
However, the magnetic moment at the Lu site could not be detected, due to noise in the diffraction profiles preventing an accurate estimation of such a small magnetic moment.\\
\indent Among the ${\bm q}=0$ structures, not only collinear structures but also noncollinear arrangements on each pyrochlore tetrahedron, such as ordered spin-ice states, are possible.
Such noncollinear configurations can generally yield detectable magnetic reflections at positions such as (200) in {\ce{Tb_2Sn_2O_7}} \cite{TbSnO1, TbSnO2}, which are absent in the collinear pyrochlore ferromagnet {\ce{Gd_2Mo_2O_7}} \cite{GMO}. However, our experiments show no significant difference in intensity at this position between $T=3$\ K and 400\ K (see the left side of Fig.\ \ref{pnd}(b)), making such a noncollinear scenario less likely.\\
%
%
%
%
%
%
%
%
%
\subsection{Single-crystal neutron diffraction measurements}
\begin{figure*}[]
\begin{tabular}{ll}
\begin{minipage}[t]{0.166\columnwidth}
    \centering
    \includegraphics[keepaspectratio, width=1\columnwidth]{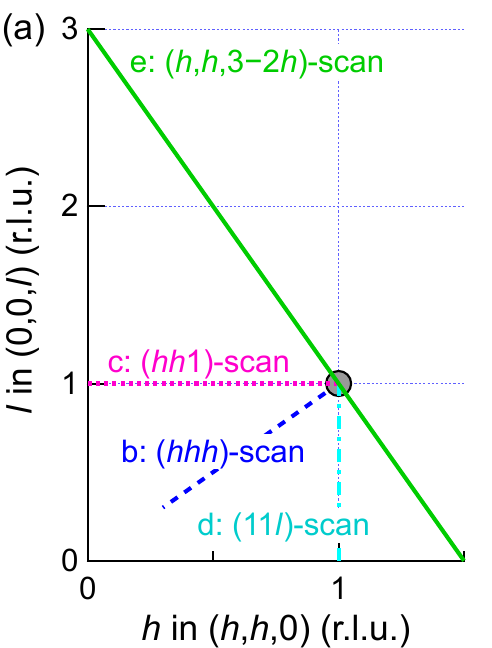}
\end{minipage}&
\begin{minipage}[t]{0.83\columnwidth}
    \centering
    \includegraphics[keepaspectratio, width=1\columnwidth]{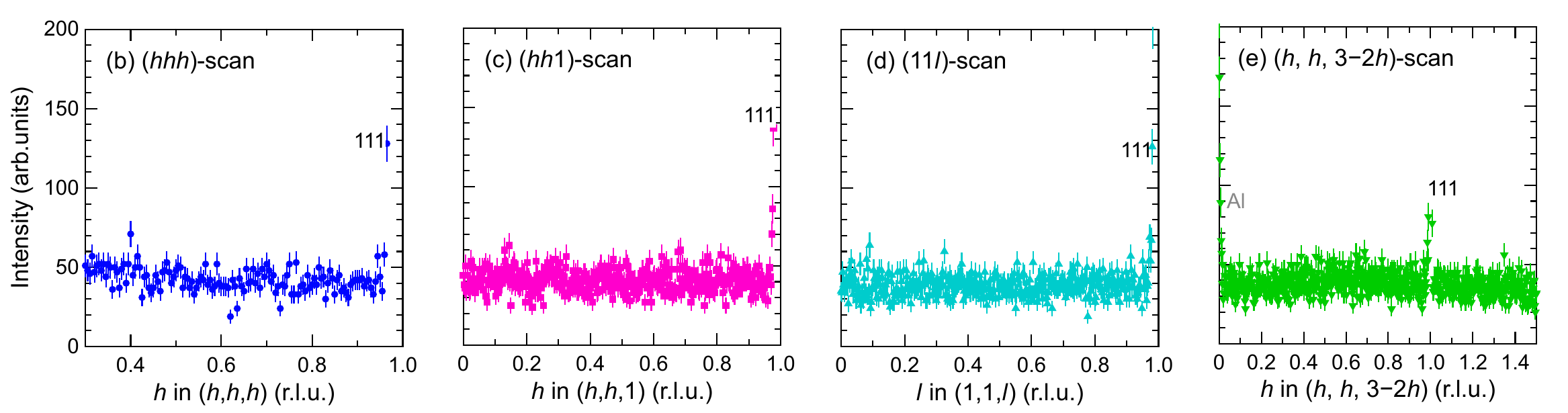}
\end{minipage}\\
\begin{minipage}[t]{0.166\columnwidth}
    \centering
    \includegraphics[keepaspectratio, width=1\columnwidth]{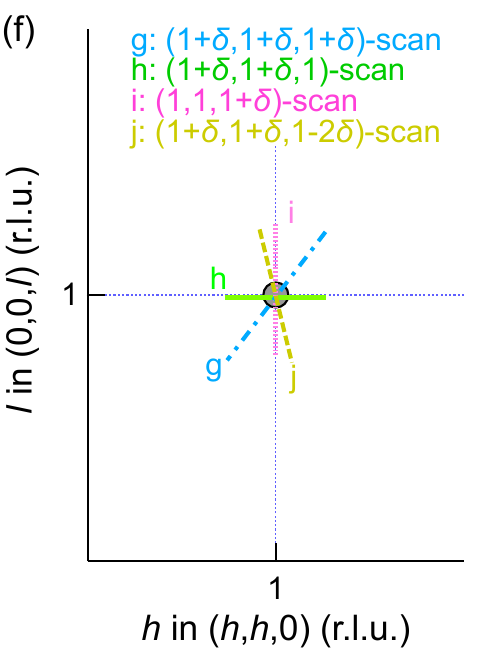}
\end{minipage}&
\begin{minipage}[t]{0.83\columnwidth}
\centering
    \includegraphics[keepaspectratio, width=1\columnwidth]{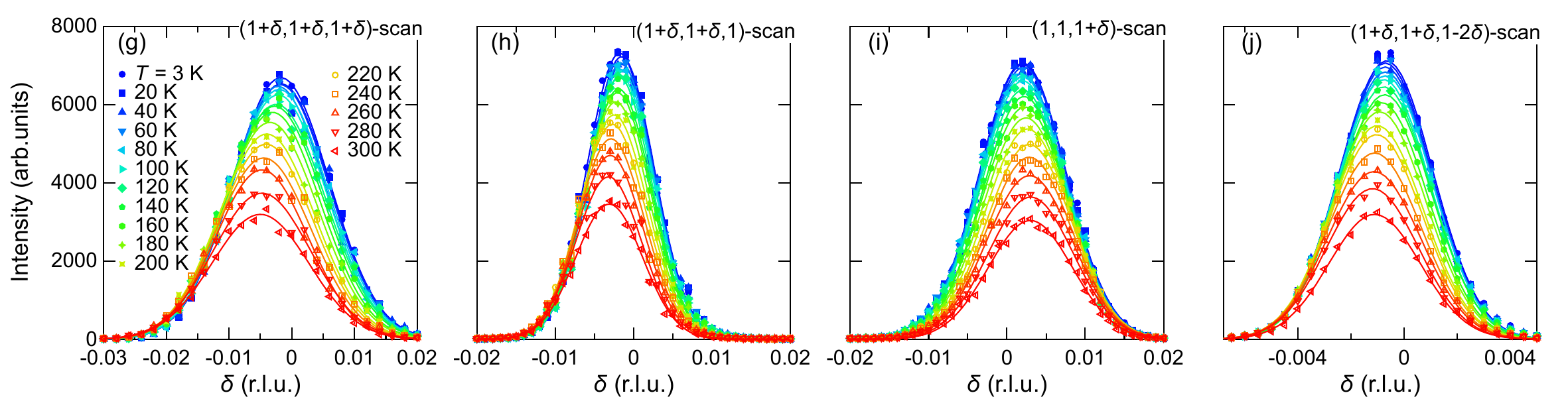}
\end{minipage}
\end{tabular}
\caption{(a), (f) Scan directions in the $(hhl)$ scattering plane for the profiles shown in (b)--(e) and (g)--(j). The circle shows the position of the magnetic and nuclear reflections at (111). (b)--(e) single-crystal neutron diffraction profiles scanned along the ($hhh$), ($hh1$), ($11l$), and ($h,h,3$$-2$$h$) lines. (g)--(j) Profiles of the (111) reflection scanned along the (1$+$$\delta$,1$+$$\delta$,1$+$$\delta$), (1$+$$\delta$,1$+$$\delta$,1), (1,1,1$+$$\delta$), and (1$+$$\delta$,1$+$$\delta$,1$-$2$\delta$) lines at different temperatures. The solid curves show the fits using a single Gaussian function.}
\label{111scan}
\end{figure*}
For a more detailed investigation of possible peak splitting or broadening related to long-period modulated magnetic structures, we measured diffraction profiles along the $(hhh)$, $(hh1)$, $(11l)$, and $(h,h,3-2h)$ directions [see Fig.\ \ref{111scan}(a)] by single-crystal neutron diffraction experiments. The results are shown in Figs.\ \ref{111scan}(b)--(e).
Apart from the nuclear reflections, no additional magnetic reflections were observed. Note that the peak at the forbidden (003) position arises from powder diffraction by the Al sample cell.\\
\begin{figure}[h]
\centering
    \includegraphics[keepaspectratio, width=0.66\columnwidth]{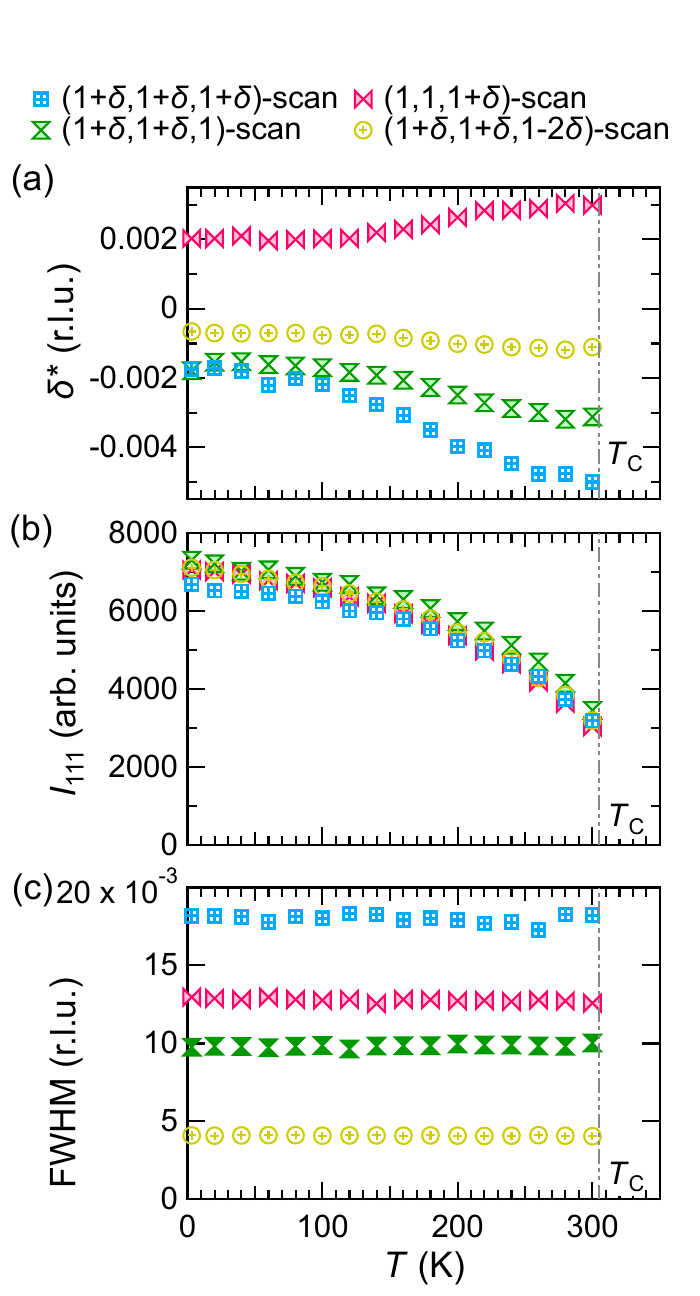}
    \caption{Temperature dependence of the peak position $\delta^{\ast}$ (a), peak height $I_{111}$ (b), and peak width FWHM (c), estimated by fitting the (111) peak profiles with a single Gaussian function.}
    \label{111gauss}
\end{figure}
\indent Next, the (111) reflection was scanned along four different lines: (1$+$$\delta$,1$+$$\delta$,1$+$$\delta$), (1$+$$\delta$,1$+$$\delta$,1), (1,1,1$+$$\delta$), and (1$+$$\delta$,1$+$$\delta$,1$-$2$\delta$) [see Fig.\ \ref{111scan}(f)]. The results are shown in Fig.\ \ref{111scan}(g)--(j). 
Upon cooling, the reflection does not show any anomalies and can be fitted by a single Gaussian function over the entire temperature range of $T = 3$--300\ K for all scan directions. The fitting parameters are shown in Fig.\ \ref{111gauss}.
While the sign of the peak shift depends on the scan direction, the peak positions $\delta^{\ast}$ change monotonically and then become constant upon cooling. This is consistent with the thermal contraction observed in the PND experiments [see Table\ \ref{nriet}].
The estimated peak heights $I_{111}$ increase monotonically upon cooling, indicating an increase in the moment size, as is typically observed in ferromagnets.
The peak widths (full width at half-maximum, FWHM) are well accounted for by the calculated instrumental resolution together with a modest sample mosaic spread. They are nearly temperature independent, indicating an absence of peak splitting associated with long-period modulation.
Together with the PND results, this suggests that {\LIC} has a ferromagnetic structure with a magnetic propagation vector ${\bm q} = 0$ over the entire temperature range.
\\
\begin{figure*}[]
\centering
    \includegraphics[keepaspectratio, width=0.9\columnwidth]{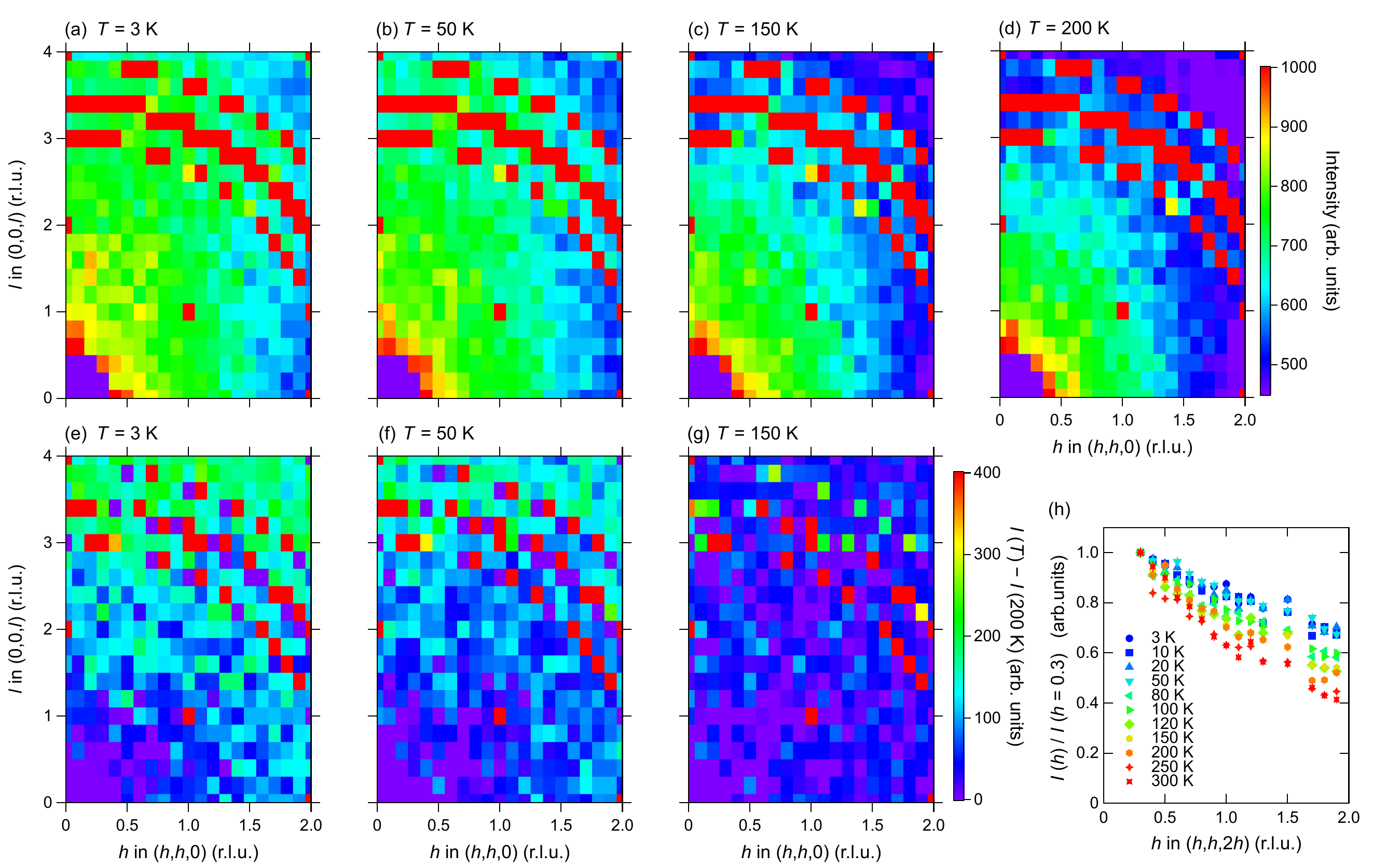}
    \caption{(a)--(d) Scattering intensity maps in the ($hhl$) scattering plane at different temperatures. (e)--(g) Difference maps relative to the scattering intensity at $T=200$ K. (h) Diffuse scattering intensity scanned along the ($h,h,2h$) direction at different temperatures. Note that the intensity at $h = 1.5$ originates from the Al powder diffraction.}
    \label{map}
\end{figure*}
\indent We recorded scattering intensity maps in the ($hhl$) scattering plane as shown in Fig.\ \ref{map}(a)--(d). Diffuse scattering was broadly distributed throughout reciprocal space.
The three ring patterns originate from Al powder diffraction peaks of the sample cell at the (111) and (200) positions and from diffraction by the stainless-steel screws that were used as the sample holder.
The diffuse scattering shows a characteristic temperature evolution: 
upon cooling, the scattering intensity extends toward higher $q$ regions.
This evolution is more clearly observed in the difference maps, which are obtained by subtracting the intensity at 200 K from the intensity at each temperature ($I(T) - I(200\ {\rm K})$), as shown in Fig.\ \ref{map}(e)--(g). We can see a more pronounced contrast in the high $q$ region at lower temperatures, while no significant difference is observed at $T = 150$\ K.
The observed diffuse scattering does not exhibit characteristic anisotropic patterns, but rather, is distributed nearly isotropically in reciprocal space. This behavior differs from the diffuse scattering observed in spin-ice systems, which reflects spatial spin correlations \cite{HTODS,HTODS2, DTODS}. Therefore, the observed scattering is more likely associated with paramagnetic spin fluctuations than with spin-ice-like correlations.
In this case, the $\bQ$ dependence of the scattering intensity is expected to be proportional to the square of the magnetic form factor of the Co--3$d$ electrons that do not participate in the ferromagnetic ordering. 
Then, the observed temperature evolution suggests that the real-space distribution of the Co--3$d$ electrons becomes extended at high temperatures and gradually localizes upon cooling. 
To investigate the temperature evolution of the diffuse scattering in detail, we performed scans along the ($h,h,2h$) direction at different temperatures. 
Since the diffuse scattering is nearly isotropic in $\bQ$ space, scans along different symmetry-equivalent directions are expected to yield equivalent results.
As shown in Fig.\ \ref{map}(h), the decay of the intensity with $h$ weakens upon cooling from $\TC$. 
Further detailed measurements on the diffuse scattering are required to clarify the microscopic origin of the gradual localization of the itinerant Co--3$d$ electrons.
\section{Discussion}
Since no additional magnetic reflections or peak splitting (broadening) were observed, it is plausible to conclude that {\LIC} has a collinear ferromagnetic structure without an antiferromagnetic component or a helical modulation.
However, magnetic diffuse scattering was clearly observed, and its temperature-dependent intensity distribution allows us to trace the spatial evolution of the magnetic Co--3$d$ electrons. The broadening of the diffuse scattering in $q$ space upon cooling can be interpreted as a broadening of the Co magnetic form factor, which corresponds to the Co--3$d$ electrons gradually becoming localized in real space. 
In itinerant-electron magnets, the difference between itinerant and localized magnetic form factors is most evident in the low $q$ region, whereas conventional magnetic Bragg reflections generally occur at higher $q$, making such a distinction difficult. 
Magnetic diffuse scattering, however, extends into the low $q$ region and is therefore sensitive to this difference. Such sensitivity has been demonstrated in LiV$_2$O$_4$, where low $q$ magnetic diffuse scattering was used to identify a partially delocalized magnetic form factor \cite{Tomiyasu}.\\
\indent The pyrochlore geometry can host flat bands arising from destructive interference of electron hopping, whose electronic states have a strong spatially localized character \cite{fb1, fb2, CaNi}. In {\LIC}, nearly flat Co--3$d$ bands are located near the Fermi level \cite{LIC}. The progressive localization inferred from the diffuse scattering may therefore reflect the development of this flat-band-derived localized character upon cooling. The magnetic anisotropy that develops below approximately 100\ K may also be associated with this evolution.\\
\indent Based on this, we propose a possible scenario for the metamagnetic transition observed below 100 K \cite{LIC}. In some itinerant-electron ferromagnets, a finite applied magnetic field can enhance the ferromagnetic moment by inducing a more strongly spin-polarized state \cite{YC3, TC3, LC3}. This behavior corresponds to an itinerant-electron metamagnetic transition, which occurs when a ferromagnetic instability persists below the magnetic ordering temperature.
Within this framework, the progressive localization of the Co--3$d$ electrons can narrow the Co--3$d$ flat bands and sharpen the peak in the density of states near the Fermi level, thereby enhancing the residual ferromagnetic instability and facilitating further spin polarization under an applied magnetic field $\Hm$.\\
\indent However, unlike conventional itinerant-electron metamagnetic systems \cite{TC3, LC3}, the transition in {\LIC} is observed only when the field is applied along the [111] direction, and the transition field $\Hm$ increases upon cooling.
These features suggest that the transition cannot be understood solely in terms of a ferromagnetic instability, and that the magnetic anisotropy developing at low temperatures may also be involved.
In a pyrochlore lattice, local magnetic anisotropy with either a $\langle111\rangle$ easy axis or a $\{111\}$ easy plane constrains the moment directions and can compete with collinear ferromagnetic order \cite{spiniceHo, YTO}. 
Such local constraints can introduce geometrical frustration and enhance magnetic fluctuations at low temperatures. Therefore, it is conceivable that these fluctuations compete with the ferromagnetic instability in {\LIC}, and that the degree of competition varies with the field direction. This could give rise to the anisotropic metamagnetic transition.
This scenario could also account for the unusual temperature dependence of the critical field. In conventional itinerant-electron metamagnetism, the temperature evolution of the critical field is determined by thermal fluctuations that are enhanced by heating. In {\LIC}, however, it may instead be determined by frustration-induced magnetic fluctuations that are enhanced by cooling.
Assessing the validity of this scenario requires an appropriate microscopic theoretical model. In addition, single-crystal magnetic diffuse scattering measurements under applied magnetic fields are necessary to clarify the role of local magnetic anisotropy.
\section{Conclusion}
We investigated the magnetic ground state of {\LIC} using powder neutron diffraction and single-crystal neutron scattering measurements under zero magnetic field.
The powder neutron diffraction data indicate that the magnetic propagation vector is ${\bm q}=0$, suggesting that the most plausible model is a collinear ferromagnetic state.
Furthermore, diffuse magnetic scattering was observed around ${\bm q}=0$, with its intensity extended toward higher $q$ regions upon cooling, suggesting a gradual localization of the itinerant Co--3$d$ electrons in real space.
This localization could enhance the residual ferromagnetic instability and, under the influence of local magnetic anisotropy on the Co-pyrochlore lattice, result in an itinerant-electron metamagnetic transition, causing the ferromagnetic moment to increase further.
These results suggest that {\LIC} is a promising platform for exploring unconventional itinerant-electron magnetism arising from local geometrical effects in pyrochlore lattices.
\begin{acknowledgments}
This work was supported by JSPS KAKENHI (Grant Nos. JP24KJ1325, JP26H02027 and JP22H05145), and JST FOREST (Grant No. JPMJFR202V).
The neutron experiments at GPTAS (4G) and HERMES were carried out by the JRR-3 general user program managed by the Institute for Solid State Physics, The University of Tokyo (Proposal Nos. 25502 and 25583, respectively).
\end{acknowledgments}

\bibliography{ref}

\end{document}